\documentclass[twocolumn,prd,amsmath,amssymb]{revtex4}
\usepackage{graphicx}
\usepackage{dcolumn}
\usepackage{bm}
\usepackage{epsfig}
\usepackage{amssymb}
\usepackage{amsmath}

\def\fun#1#2{\lower3.6pt\vbox{\baselineskip0pt\lineskip.9pt
  \ialign{$\mathsurround=0pt#1\hfil##\hfil$\crcr#2\crcr\sim\crcr}}}

\newcommand{\PRD}{Phys.\ Rev.\ D.} 
\newcommand{\PRL}{Phys.\ Rev.\ Lett.}

\def\fun#1#2{\lower3.6pt\vbox{\baselineskip0pt\lineskip.9pt
\ialign{$\mathsurround=0pt#1\hfil##\hfil$\crcr#2\crcr\sim\crcr}}}

\def\mpl{m_{\rm Pl}}
\newcommand{\MUNCH}[1]{\relax}

\begin{document}

\title{Deciphering Inflation with Gravitational Waves: Cosmic Microwave Background Polarization vs. Direct Detection with Laser Interferometers}

\author{Tristan~L.~Smith$^1$, Hiranya~V.~Peiris$^{2}\footnote{Hubble Fellow, {\sf hiranya@cfcp.uchicago.edu}}$, Asantha~Cooray$^3$}
\affiliation{$^1$California Institute of Technology, Mail Code 130-33, Pasadena, CA~91125\\
$^2$Kavli Institute for Cosmological Physics and Enrico Fermi Institute, University of Chicago, Chicago IL~60637\\
$^3$Center for Cosmology, Department of Physics and Astronomy, University of California, Irvine, CA~92697}

\date{\today}

\begin{abstract}

A detection of the primordial gravitational wave background is considered to be the ``smoking-gun'' evidence for inflation. While super-horizon waves are probed with cosmic microwave background (CMB) polarization, the relic background will be studied with laser interferometers. The long lever arm spanned by the two techniques improves constraints on the inflationary potential and validation of consistency relations expected under inflation. If gravitational waves with a tensor-to-scalar amplitude ratio greater than 0.01 are detected by the CMB, then a direct detection experiment with a sensitivity consistent with current concept studies should be pursued vigorously. If no primordial tensors are detected by the CMB, a direct detection  experiment to understand the simplest form of inflation must have a sensitivity improved by two to three orders of magnitude over current plans.
\end{abstract}
   
\maketitle

\section{Introduction}

Recent high precision measurements of the cosmic microwave background
(CMB) anisotropy power spectrum \cite{Spergel:03} have confirmed inflation \cite{Guth81}
as the dominant paradigm to explain the origin of primordial fluctuations with a nearly scale-invariant spectrum. In addition to density perturbations, inflationary models predict a stochastic background of gravitational waves \cite{Abbott84,Star79} with the amplitude of the gravitational wave background given by the height of the inflaton potential when relevant modes exit the horizon.

Inflationary gravitational waves with wavelengths comparable to the horizon size are now being sought via ground-based CMB polarization experiments \cite{Kamionkowski:96}, and eventually 
with a dedicated satellite generally named CMBPol \cite{CMBPol}. In addition to the CMB effort, 
concept studies are now underway to investigate the possibility of directly detecting the relic background 
with a laser interferometer in space (Big Bang Observer \cite{BBO}; for the DECIGO proposal in Japan, see Ref.~\cite{Seto:01}). The direct detection technique will be sensitive to modes with wavelengths roughly an arm length of the interferometer. Based on the expected foreground confusion and technological improvements, current concept studies aim for the frequency regime between 0.1 Hz to a few Hz. Since physical scales probed by the CMB and laser interferometers  differ by $\sim$17 orders of magnitude in wavelength, the large lever arm produced by combining the two techniques allows the inflaton potential to be pinned down better than any single method \cite{Turner:96,Ungarelli:05,Smith:06,Cooray:05}.

Here, we consider two separate studies on inflation using the combined information from CMB and direct detection experiments. First,
based on a Monte Carlo description of the inflationary dynamics, we study the relative abilities of the CMB and a direct detection method to probe the inflaton potential in detail by making no assumptions on the power-law behavior \cite{Ungarelli:05} or on a model shape for the potential \cite{Smith:06}.
We first consider inflationary models allowed by CMB data, making use of constraints on both the scalar and tensor spectrum, and address if laser interferometers can further improve the identification of potentials. For the CMB, we make use of the expected level of uncertainty with Planck \footnote{http://www.rssd.esa.int/SA/PLANCK/docs/Bluebook-ESA-SCI(2005)1.pdf}, a possible detection with CMBPol, and the foreground-limit from CMBPol. For details on the potential detectability of the tensor amplitude with CMB polarization observations, we refer the
reader to Ref.~\cite{verde/etal:2005}. For the direct detection experiments, we make use of predictions related to BBO and DECIGO \cite{Seto:06}. 
Since our goal is to see how the simplest models of inflation can be constrained, our comparisons are for a general single-field potential in the slow-roll regime. 

In the second part of the paper, we drop all  assumptions related to single-field slow-roll inflation and study how well 
the CMB and the direct detection experiments can be combined to constrain the single-field consistency relation between tensor spectral index and the ratio of tensor-to-scalar  
amplitudes for single-field inflationary models. Previous work on this possibility is found in Ref.~\cite{Song} where the test was limited to simply the information on the tensor spectral index from CMB data alone.
We show the extent to which a CMB-only analysis can be improved by adding direct-detection information as the latter allows a better determination of the tensor spectral index.

This paper is organized as follows: In the next Section, we present the Monte Carlo modeling of the inflationary potential that satisfy both detections
or limits at CMB and laser interferometer scales. In Section~III, we discuss a fundamental test of single-field slow-roll inflation, involving
the consistency relation between tensor-to-scalar ratio and the tensor spectral index. We conclude with a brief summary of our results in
Section~IV. 

\section{Monte Carlo Calculational Method}

To calculate observable spectra, we make use of a Monte Carlo technique
to formulate the inflationary dynamics through an infinite hierarchy of flow equations involving the generalized ``Hubble Slow Roll'' (HSR) parameters \citep{Schwarz,hoffman/turner:2001, kinney:2002,
easther/kinney:2003}. We link the Hubble parameter directly to the field $\phi$ instead of
time, $H \equiv H(\phi)$, under the assumption that $\phi$ is monotonic in time.
The equation of motion for the background is
\begin{eqnarray}
\left[H'(\phi)\right]^2 -
\frac{12\pi}{\mpl^2}H^2(\phi)&=&-\frac{32\pi^2}{\mpl^4}V(\phi) \, , \label{eq:hj}
\end{eqnarray}
where the inflaton field  evolves as
$\dot{\phi} =- \mpl^2 H'(\phi)/4\pi$.
Here, an overdot corresponds to the time derivative and a prime denotes the derivative with respect to $\phi$. 
The advantage of this formulation is that one can study the  generic behavior of slow roll single field inflation without assuming a particular shape for the potential, except for the assumption of a single field. In terms of the HSR parameters $^{\ell}\lambda_H$, the dynamics of inflation is described through:
\begin{eqnarray}
\epsilon(\phi) &\equiv& \frac{m^2_{\rm Pl}}{4\pi}
\left[\frac{H'(\phi)}{H(\phi)}\right]^2; \label{eq:eps} \\
^{\ell}\lambda_H &\equiv& \left(\frac{m^2_{\rm Pl}}{4\pi}\right)^\ell
  \frac{(H')^{\ell-1}}{H^\ell} \frac{d^{(\ell+1)} H}{d\phi^{(\ell+1)}}
   ;\ \ell \geq 1.  \label{eq:hier}
\end{eqnarray}
Substituting equation~(\ref{eq:hier}) in equation~(\ref{eq:hj}) gives
the inflaton potential 
\begin{equation}
V(\phi) =
\left(\frac{3\mpl^2 H^2\left(\phi\right)}{8\pi}\right)
\left[1-\frac{1}{3}\epsilon\left(\phi\right)\right] \, . \label{eq:v}
\end{equation}

The trajectories of the flow parameters are now governed by a set of coupled first order differential equations. In practice, one has to truncate the infinite hierarchy at some finite order; in this paper we retain terms up to 10th order.  Truncating the hierarchy of flow parameters at the term $^M\lambda_H$ means that $^{M+1}\lambda_H = 0$ at all times as well. From Eq.~\ref{eq:hier}, it also follows that $d^{(M+2)} H/d\phi^{(M+2)} = 0$ at all times. This simply describes a polynomial of order $M+1$ in $H(\phi)$ \citep{liddle:2003} with
\begin{equation}
H(\phi) = H_0\left[ 1+ A_1 \left(\frac{\phi}{\mpl}\right) + ... + A_{M+1}
\left(\frac{\phi}{\mpl}\right)^{M+1}\right]. \label{eq:h}
\end{equation}
Further, from the definition of $\epsilon(\phi)$, 
\begin{eqnarray}
&&\epsilon(\phi) = \frac{m^2_{\rm Pl}}{4\pi} \\
&\times&\left[\frac{\left(A_1/\mpl\right) + ... + (M+1)
\left(A_{M+1}/\mpl\right)\left(\phi/\mpl\right)^M}{1+A_1\left(\phi/\mpl\right)
+ ... + A_{M+1}\left(\phi/\mpl\right)^{M+1}}\right]^2, \nonumber  \label{eq:epsanalytic}
\end{eqnarray}
when the coefficients $A_i$, with $i > 1$, are written in terms of the initial values
of the HSR parameters as
\begin{eqnarray}
A_{\ell+1} &=& \frac{(4\pi)^\ell \ ^{\ell}\lambda_{H,0}}{(\ell+1)! 
\ A_1^{\ell-1}} \, , \label{eq:coeffs}
\end{eqnarray}
where $ A_1 = \sqrt{4\pi\epsilon_0}$ specifies the direction the field is rolling.  These slow roll parameters require a prior assumption on the ranges of values taken. In the absence of any {\sl a priori} theoretical knowledge, one can assume flat priors with the requirement that the potential satisfies
the slow-roll condition; the latter is simply a statement about the smoothness of the potential. Since we are not attempting to make any statements about the measure of inflationary trajectories, but simply use the method as a Monte Carlo generator for potentials satisfying the slow roll conditions, such an
assumption is justified. This work does not depend on the density of potentials as a function of tensor-to-scalar ratio, which is determined by the measure on the initial conditions.

\begin{figure*}[!ht]
\includegraphics[scale=1.]{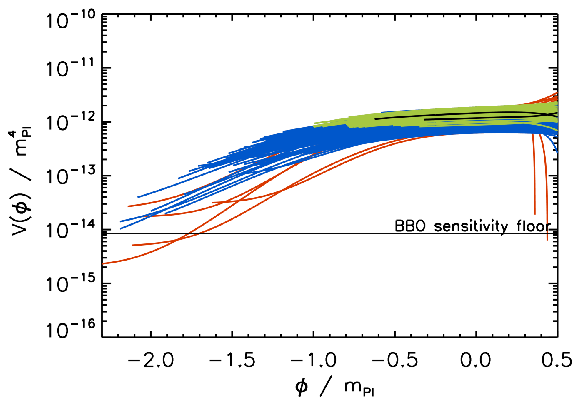}
\includegraphics[scale=1.]{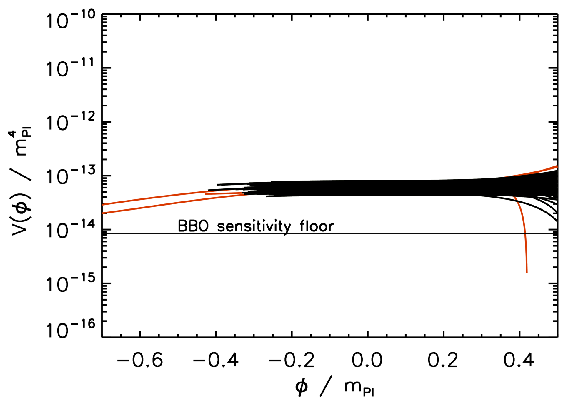}
\includegraphics[scale=1.]{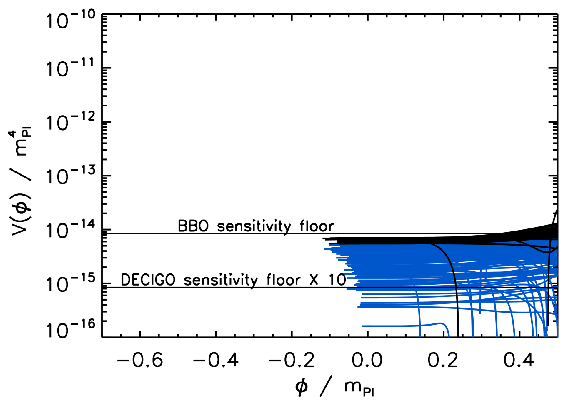}     
\caption{\small The set of potentials satisfying $0.9<n_s<1.0$ from the Monte Carlo flow simulations:
{\it Left}: constraints from Planck with $r=0.02 \pm 0.01$, $d n_s/d\ln k = 0.0\pm 0.01$, BBO-standard, BBO-grand (factor 10 more sensitive than BBO-standard) {\it Center}: constraints from CMBPol with optimistic foregrounds \cite{verde/etal:2005}: $r=0.001 \pm 0.0003$, $d n_s/d\ln k = 0.0\pm 0.005$, BBO-standard, BBO-grand, and {\it Right}: the sensitivity limit due to foregrounds for CMBPol with $r<10^{-4}$ and DECIGO (factor 400 more sensitive than BBO-standard). {\it Color coding}: red and blue denote the CMB experiment without and with the $d n_s/d\ln k$ constraint, respectively. Green: BBO-standard. Black: BBO-grand ({\it Left, Center}) and DECIGO ({\it Right}). The direct detection constraints are applied to the tensor amplitude and tilt at 1 Hz, following the procedure described in the text. 
If a particular color does not appear in a plot, it has been overwritten by the next tightest constraint, and the latter is therefore {\it not} helpful in constraining the potential. The meaning of the color coding is further clarified in the text. Here $\phi=0$ corresponds to CMB scales while curves end at $\phi <0$ corresponding to a frequency of 1 Hz probed by direct detection methods.}
\label{fig:cmbrlimits}
\end{figure*}

Without loss of generality, we can pick some fiducial physical scale that corresponds to $\phi_{CMB}$, which we take to be at $k_{\rm CMB}=0.002$ Mpc$^{-1}$. 
Then, with the above convention, $\phi>\phi_{CMB}$
corresponds to scales larger than $k_{\rm CMB}$ (i.e. going further back in time), and $\phi<\phi_{CMB}$ corresponds to smaller scales. The physical wavenumber is associated with a value of $\phi$ through
\begin{equation}
\frac{d\phi}{d\ln k} = -\frac{\mpl}{2\sqrt{\pi}}
\frac{\sqrt{\epsilon}}{1-\epsilon}\, , \label{eq:phieq}
\end{equation}
while the number of e-folds before the end of inflation, $N$, comes from
$d\phi/dN = \mpl\sqrt{\epsilon}/2\sqrt{\pi}$,
with the convention that $N$ increases as one goes further back in time. Here, we require that each
potential generated by the Monte Carlo flows provide at least $N=55$ e-folds of inflation. For potentials where inflation ends through the breakdown of slow-roll, the CMB observables are calculated at 55 e-folds before the end of inflation; for potentials corresponding to the hybrid case, the CMB observables are arbitrarily calculated at the 600th e-fold, assuming inflation ends at 655 e-folds through an orthogonal mechanism.

The standard observables are given in terms of the flow parameters to second order in the slow roll \cite{stewart/lyth:1993,liddle/etal:1994}:
\begin{eqnarray}
n_s &=& 1 + 2\eta - 4\epsilon - 2(1+C) \epsilon^2 \label{eq:hsrconversionsstart} \\
&-& \frac{(3-5C)}{2} \epsilon \eta + \frac{(3-C)}{2}\xi \nonumber \\  
r &=& 16 \epsilon \left[1+2C(\epsilon - \eta)\right] \label{eq:r} \\
n_t &=& -2\epsilon - (3+C) \epsilon^2 + (1+C) \epsilon \eta \, , \label{eq:nt}
\end{eqnarray}
where $n_s$, $r$, and $n_t$ are the tilt of the scalar spectrum, tensor-to-scalar ratio, and the tilt of the tensor spectrum, respectively. Additionally, we consider the running of the scalar tilt with
\begin{eqnarray}
\frac{dn_s}{d \ln k} &=& -\frac{1}{1-\epsilon} \Big\{2 \frac{d\eta}{dN} - \left[4 + 4\left(1+C\right) \epsilon\right]\frac{d\epsilon}{dN}  \\
&-& \frac{(3-5C)}{2}\left(\epsilon\frac{d\eta}{dN} + \eta \frac{d\epsilon}{dN}\right) + \frac{(3-C)}{2}\frac{d\xi}{dN} \Big\} \, , \nonumber
\end{eqnarray}
where $d\epsilon/dN = 2\epsilon(\eta-\epsilon)$, $d\eta/dN = -\epsilon\eta + \xi$, and
$d\xi/dN = \xi (\eta-2\epsilon) +\  ^3\lambda_H$. Here and above,
$C=4(\ln 2 + \gamma) - 5$ and $\gamma=0.5772$ is the
Euler-Mascheroni Constant, $\eta =\ ^1\lambda_H$, and $\xi =\
^2\lambda_H$. 

For direct detection experiments, we take a fiducial frequency of  $f = 1$ Hz with $k_{\rm dir} = 6.47 \times 10^{14}$ Mpc$^{-1}$ for observations. With $\ln(k_{\rm dir}/k_{\rm CMB}) = 40.3$ the large lever arm is expected to improve constraints on the inflationary model \cite{Smith:06}.  The direct detection observables are calculated from the potentials by finding the $\phi$ corresponding to $\ln(k_{\rm dir}/k_{\rm CMB}) = 40.3$. In these experiments, at 1 Hz, the signal-to-noise ratio for a detection of the gravitational wave background is ${\rm SNR} = X (\Omega_{\rm GW}/10^{-18})$, where $X$ is $\sim$ 0.25, 2.5, and 100 for concept study designs involving a standard BBO, an optimistic version of BBO,
and DECIGO. The last two possibilities improve sensitivities through multiple detector correlations. 
The uncertainty of the tensor spectral index at 1 Hz is taken to be $\sigma n_T \sim 6/{\rm SNR}$ \cite{Seto:06}. 

Fig.~\ref{fig:cmbrlimits} shows a set of potentials from the Monte Carlo flow simulations that satisfy levels of tensor and scalar modes reachable by Planck,  a CMBPol-style polarization satellite designed to probe primordial gravity waves \cite{verde/etal:2005}, and the foreground limit of this satellite. We show all potentials with $0.9 < n_s < 1.0$, though in practice the CMB measurements will also yield strong constraints on this parameter. In Fig.~1 we also show the improvement in constraints on the potential using information from direct detection experiments with uncertainties in the tensor amplitude and tilt calculated following results of the analysis in Ref.~\cite{Seto:06}. 

In Fig.~1, the potentials generated by the Monte-Carlo process are color-coded in the following way. First, all potentials with $0.9 < n_s < 1.0$ which are {\it not} ruled out by applying the constraints on $r$ from the CMB experiment  are shown in red. Then the potentials which are {\it not} ruled out by further applying the $d n_s/ d \ln k$  constraint from the CMB experiment are over-plotted in blue. Next, the potentials which are {\it not} eliminated by adding the constraint from BBO-standard on the tensor amplitude and tilt to the CMB experiment are over-plotted in green. Finally, the potentials which are {\it not} eliminated by adding the constraints from either BBO-grand (left, centre) or DECIGO (right) to the CMB experiment are over-plotted in black. At each stage, the elimination of a larger fraction of potentials indicates the usefulness of the extra constraints. If a particular color does not appear in a given panel, it means that all potentials have been overwritten by the next tightest constraint, and therefore the next tightest constraint does not aid in constraining the potential. Therefore, while the combination of Planck and BBO-standard, and especially BBO-grand, leads to an improvement in constraints on the potential, these direct detection experiments do not compete at all with CMBPol either in the case of a detection at $r=0.001$ or the limit of $r < 10^{-4}$ at CMB scales.  A more sensitive experiment such as DECIGO probing $\Omega_{\rm GW}h^2 > 10^{-20}$  competes well with CMBPol if $r \lesssim 0.01$ and is both desirable and useful to understand inflation.

\section{A fundamental test of slow-roll Inflation}

\begin{figure*}[!ht]
\includegraphics[scale=0.6]{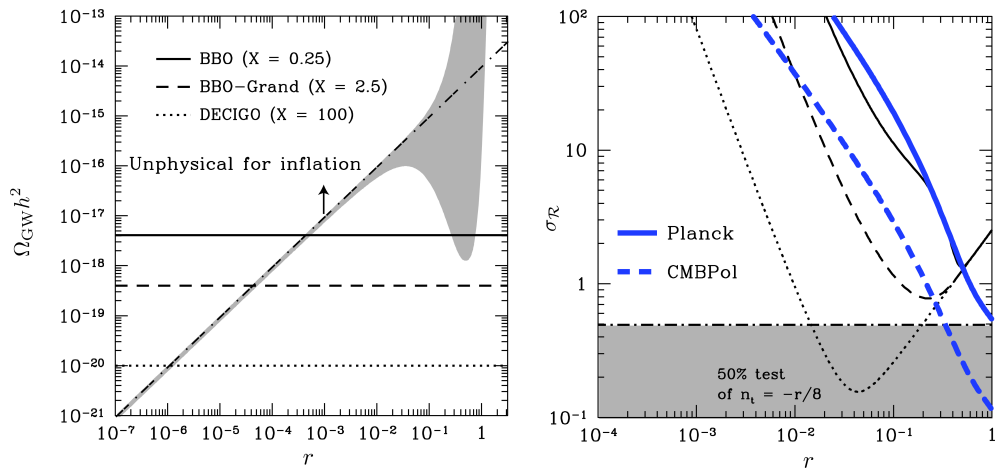}
\caption{{\it Left:} 
The mapping between tensor-to-scalar ratio $r$ at CMB scales and $\Omega_{\rm GW}h^2$
at 1 Hz for a direct detection. The gray shaded region shows the uncertainty implied with $n_s=0.95 \pm 0.1$
\cite{Seljak} by keeping terms up to second order (running of the tensor spectral index) in the slow-roll power-law expansion.  We include the line `Unphysical for inflation' to indicate the region above which $n_t > 0$.
The three horizontal curves are the 2$\sigma$ detection amplitudes for
 BBO-standard, BBO-grand, and DECIGO in solid, dashed, and dotted lines respectively.
In these three experiments, at 1 Hz, the signal-to-noise ratio for detecting the gravitational wave background is 
${\rm SNR} = X (\Omega_{\rm GW}/10^{-18})$, with values for $X$ shown in the panel.
{\it Right:} The $1 \sigma$ uncertainty in the single-field consistency relation
$\mathcal{R}\equiv-r/8n_t$.  The thin lines follow laser interferometers
in the right panel, showing the error expected by combining direct detection measurements of $n_t$ with
CMBPol measurements of the tensor-to-scalar ratio, $r$.  The two thick lines indicate errors in $\mathcal{R}$ when $n_t$ is determined from CMB alone.  The thick solid curve corresponds to the expected accuracy of ESA's Planck satellite, the thick dashed curve corresponds to CMBPol.  The shaded region indicates a 50\% determination of the consistency relation ($\mathcal{R} = 1.0 \pm 0.5$).  For direct-detection observations, sensitivity is degraded at large $r$ because of an increase in the uncertainty of the importance of the running of the tensor spectral index from direct detection scales to CMB scales; at small $r$, the accuracy with which $\mathcal{R}$ can be determined is dominated by the error in measuring $n_t$ with CMB \cite{Song} and laser interferometers \cite{Seto:06}, where the latter is
$\sigma_{\small n_t} \propto 1/r$.
}
\label{fig:consistency}
\end{figure*}

To further understand the usefulness of a direct detection experiment, we also study how well the inflationary single-field consistency condition can be tested \cite{Song}, as any departure can capture important physics \cite{Kaloper}.  We define $\mathcal{R} \equiv -r/8 n_t$ so that $\mathcal{R} = 1$ corresponds to the consistency relation.  The uncertainty in determining $\mathcal{R}$ (given $\mathcal{R} = 1$ as the fiducial value) can be written in terms of uncertainty in measurements of $r$ and $n_t$ assuming these are uncorrelated, 
\begin{equation}
\sigma_{\mathcal{R}} = \left[ \sigma_{r}^2 + 64 \sigma_{n_t}^2 \right]^{1/2} r^{-1}. \label{eq:consistency_error}
\end{equation}
A 10\% determination of $\mathcal{R}$ with the expected value of unity requires measuring $n_t$ with an uncertainty of $\sigma_{n_t} \sim 0.0125r$. Using tables of Ref.~\cite{verde/etal:2005}, CMBPol with a 10\% foreground contamination and $r=0.01$ gives $\mathcal{R}=1.0 \pm 80$. The obstacle is the inability of CMB polarization observations to measure $n_t$ precisely, since polarization anisotropies probe a limited  range in the underlying tensor spectrum modes, and the range probed is also contaminated  by cosmic shear \cite{Kesden}.  

When combining with a direct detection experiment, however, the situation improves significantly. We calculate the 1$\sigma$ error expected in $\mathcal{R}$ at CMB scales based on a determination of $n_t$ from direct detection scales and a determination of $r$ by CMBPol.  In Ref.~\cite{Seto:06} it was shown that a laser interferometer (whose sensitivity peaks at $\sim 1$ Hz) would be able to determine the spectral tilt of a gravitational wave background, with an amplitude $\Omega_{\rm GW}$, to an accuracy of 
\begin{equation}
\sigma_{n_t} = \frac{6 \times 10^{-18}}{X \Omega_{\rm GW} h^2},
\end{equation}
where the various values of $X$ corresponding BBO-standard, BBO-grand and DECIGO are shown in Fig.~2 and $h$ is the Hubble parameter today in units of $100\ \mathrm{km}\,\mathrm{s}^{-1} \,
\mathrm{Mpc}^{-1}$.  

While we take our 
fiducial model to be single-field slow -roll inflation (i.e. $\mathcal{R}=1.0$), we make no assumptions about single-field slow-roll inflation when we relate $\Omega_{\rm GW} h^2$ to the tensor-to-scalar ratio.  First we relate $\Omega_{\mathrm{GW}}h^2$ to the primordial spectrum as in Ref.~\cite{Smith:06}, 
\begin{equation}
     \Omega_{\mathrm{GW}}(k) h^2 = A_{\mathrm{GW}} P_t(k),
\end{equation}
where $P_t(k)$ is the primordial power spectrum of inflationary gravitational waves, and $A_{\mathrm{GW}} = 2.74 \times 10^{-6}$.  The factor, $A_{\mathrm{GW}}$, takes into account how the gravitational waves have evolved after re-entering the horizon. 
The primordial inflationary gravitational wave power spectrum can be approximated as a power law with a running spectral index, 
\begin{equation}
P_t(k) \approx P_t(k_0) \left(\frac{k}{k_0}\right)^{n_t + \frac{1}{2} \alpha_t \ln(k/k_0)}.
\end{equation}
Using the expressions in Eqs.~(\ref{eq:hsrconversionsstart}-\ref{eq:nt}) to first order in the slow roll parameters along with the expression for the running of the tensor spectral index, $\alpha_t$,  
\begin{eqnarray}
     \alpha_t(k) &\simeq& 4 \epsilon \eta - 8 \epsilon^2 \label{eq:running},
\end{eqnarray}
we are able to express $n_t$ and $\alpha_t$ in terms of $r$ and $n_s$, 
\begin{equation}
P_t(k) \approx r P_s(k_0) \left(\frac{k}{k_0}\right)^{-\frac{r}{8}\left[1-\left((n_s -1)+\frac{r}{8}\right)\ln(k/k_0)\right]}, \label{eq:Pt}
\end{equation}
where $P_s(k_0)$ is the amplitude of the scalar perturbations at some pivot wavenumber $k_0$ and all of the spectral quantities are also measured at this wavenumber.  For this analysis we take $n_s = 0.95$ and $P_s(k_0) = 2.21 \times 10^{-9}$ at $k=0.002\ \mathrm{Mpc}^{-1}$ \cite{Spergel:03}. 
  
Since we must connect a measurement of $n_t$ at BBO scales to the $n_t$ at CMB scales in order to determine $\mathcal{R}$, we must posit some scale-dependent relation between these two measurements.  We emphasize that we \emph{cannot} use the flow equation approach of the previous section here; those equations assume single-field inflation and therefore implicitly embody the consistency relation, whereas here we are attempting to {\it test} the single-field assumption.  Instead, we assume that the gravitational wave spectrum is close to scale invariant with an unknown, but higher order, running so that $n_t^{\mathrm{BBO}} \approx n_t^{\mathrm{CMB}}$.  Since the running of $n_t$ is unlikely to be determined by either the CMB or direct detection we include an additional uncertainty due to the unknown running, $\sim (1/2)\ln(k_{\mathrm{dir}}/k_{\mathrm{CMB}}) n_t^2$, as in Eq.~(\ref{eq:running}).  This leads to a decrease in the sensitivity of direct detection measurements of $\mathcal{R}$ at `large' values of $r$.  In particular, the error on our determination of $n_t$ is given by the expression, 
\begin{equation}
\sigma_{n_t}(r) = \Bigg\{\left[\frac{6 \times 10^{-18}}{X A_{\mathrm{GW}} P_t(k)}\right]^2 + \left[\frac{1}{2}\ln\left(\frac{k_{\mathrm{dir}}}{k_{\mathrm{CMB}}}\right) \left(\frac{r}{8}\right)^2\right]^2\Bigg\}^{1/2}.
\end{equation}

Our results are summarized in Fig.~2. The left panel shows the mapping between tensor-to-scalar ratio $r$ at CMB scales and $\Omega_{\rm GW}h^2$
at 1 Hz for a direct detection. After this paper appeared in preprint, another study \cite{Efs} was submitted, where a criticism was made of a region on this panel, describing it as unphysical. In fact this region is {\it only} unphysical if inflation is being assumed as the generating mechanism for tensor modes, and one cannot do that if one is attempting, as we are, to test that assumption in the first place. Therefore we continue to show this region, while indicating that it is unphysical under the assumption of inflation. As shown in the right panel, BBO improves relative to CMBPol alone by a factor of $\sim$ a few in the uncertainty of $\mathcal{R}$ if $r \sim 0.1$. This is unlikely to be useful given that current observations already limit $r$ to be below 0.3. A determination of $\mathcal{R}$  as $1.0 \pm 0.5$ is achievable when $10^{-1} \gtrsim r \gtrsim 10^{-2}$  with DECIGO, while if $r \sim 10^{-4}$, it is unlikely that even DECIGO would provide a determination of $\mathcal{R}$ to a reasonable accuracy. In general, either version of BBO is unlikely to be competitive with CMBPol, and the sensitivity level of DECIGO must be considered as the experimental target goal to pursue a direct detection at $\sim$ 1 Hz. Further limits on $r$ from the CMB will tighten this conclusion and could only lead to a further increase in required sensitivity unless a detection is made with the CMB at $r > 0.01$. Our conclusions are independent of the choice of a fiducial frequency between 0.1 Hz and few Hz, but could be subjected to the highly uncertain impact of foregrounds at direct detection frequencies \cite{Seto:06}.
Additional physics between the CMB and 1 Hz scales \cite{Boyle:05} 
only strengthen our conclusions on the
required sensitivity for a laser interferometer as these exotic models generally lower the tensor amplitude further.

\section{Summary}

To summarize, we
 have considered the relative strengths of CMB polarization observations and direct detection laser interferometers in constraining the inflaton potential.  For single-field slow-roll inflation models, without relying on any particular shape for the potential, we find that direct detection experiments with sensitivities around BBO can improve constraints on inflationary models relative to Planck. However, when combined with CMBPol, these direct detection sensitivities are unlikely to be competitive. While we have not considered exotic models, and a case can certainly be made for a low-sensitivity direct detection experiment based on non-standard descriptions for inflation including  models of bubble nucleation \cite{NS} and pre-Big Bang descriptions \cite{preb}, 
it is also important to understand first how these experiments test the simplest forms of inflation. 

In this context, we also discuss a determination of the single-field slow-roll consistency relation, which is a way to establish an underlying model within the inflationary paradigm to probe physics at the earliest times of the Universe. In general, we find an experiment like DECIGO, with sensitivity level  of $\Omega_{\rm GW}h^2 > 10^{-20}$ to be the preferred option; however, it is unlikely that the consistency relation will be determined to the accuracy needed to see loop corrections \cite{Kaloper}, unless the tensor-to-scalar ratio is greater than 0.05 and loop corrections are at the level of 10\% or more.

\acknowledgments{After completing this paper, we became aware of the preprint \cite{Efs} on the same topic using an approach similar to ours;
we thank George Efstathiou for useful discussions on their calculation.
We thank Richard Easther and Naoki Seto for useful discussions. HVP is supported by NASA through Hubble Fellowship grant \#HF-01177.01-A awarded by the Space Telescope Science Institute.  TLS acknowledges support from the NSF. AC is supported by the DOE at UC Irvine.}

\end{document}